\documentclass[useAMS,usenatbib]{mn2e}

\renewcommand\sfdefault{ptm}
\usepackage{sfmath}

\usepackage{lscape}
\usepackage{aas_macros}

\newbox\grsign \setbox\grsign=\hbox{$>$} \newdimen\grdimen \grdimen=\ht\grsign
\newbox\simlessbox \newbox\simgreatbox \newbox\simpropbox \newbox\wtildebox 
\setbox\simgreatbox=\hbox{\raise.5ex\hbox{$>$}\llap
    {\lower.5ex\hbox{$\sim$}}}\ht1=\grdimen\dp1=0pt
\setbox\simlessbox=\hbox{\raise.5ex\hbox{$<$}\llap
    {\lower.5ex\hbox{$\sim$}}}\ht2=\grdimen\dp2=0pt

\newcommand{\msun}{\mbox{${\rm M}_{\odot}$}}

\newcommand{\be}{\mbox{\begin{equation}}}
\newcommand{\ee}{\mbox{\end{equation}}}

\newcommand{\Cref}{\mbox{$m_{\rm ref}$}}

\title{The dynamical state of stellar structure in star-forming regions}   
\author{J.~M.~Diederik Kruijssen,$^{1,2,3}$\thanks{kruijssen@mpa-garching.mpg.de} Thomas Maschberger,$^{4,5,6}$ Nickolas Moeckel,$^4$ \newauthor  Cathie~J.~Clarke,$^4$ Nate~Bastian$^{4,7}$ and Ian~A.~Bonnell$^8$\\
$^{1}$Astronomical Institute, Utrecht University, PO Box 80000, 3508 TA Utrecht, the Netherlands\\
$^{2}$Leiden Observatory, Leiden University, PO Box 9513, 2300 RA Leiden, the Netherlands\\
$^{3}$Max-Planck Institut f\"{u}r Astrophysik, Karl-Schwarzschild-Stra\ss e 1, 85748 Garching, Germany\\
$^{4}$Institute of Astronomy, University of Cambridge, Madingley Road, Cambridge CB3 0HA, United Kingdom\\
$^{5}$Argelander-Institut f\"{u}r Astronomie, Auf dem H\"{u}gel 71, 53121 Bonn, Germany\\
$^{6}$UJF-Grenoble 1 / CNRS-INSU, Institut de Plan\'{e}tologie et d`Astrophysique de Grenoble, UMR 5274, Grenoble, 38041, France\\
$^{7}$Excellence Cluster Universe, Boltzmannstra\ss e 2, 85748 Garching, Germany\\
$^{8}$Scottish Universities Physics Alliance (SUPA), School of Physics and Astronomy, University of St. Andrews, North Haugh, St. Andrews, Fife KY16 9SS, \\United Kingdom}

\begin{document}
\date{Accepted 2011 September 1. Received 2011 August 17; in original form 2011 April 19}

\pagerange{\pageref{firstpage}--\pageref{lastpage}} \pubyear{2011}
\label{firstpage}

\maketitle

\begin{abstract}
The fraction of star formation that results in bound star clusters is influenced by the density spectrum in which stars are formed and by the response of the stellar structure to gas expulsion. We analyse hydrodynamical simulations of turbulent fragmentation in star-forming regions to assess the dynamical properties of the resulting population of stars and (sub)clusters. Stellar subclusters are identified using a minimum spanning tree algorithm. When considering only the gravitational potential of the stars and ignoring the gas, we find that the identified subclusters are close to virial equilibrium (the typical virial ratio $Q_{\rm vir}\approx 0.59$, where virial equilibrium would be $Q_{\rm vir}\sim 0.5$). This virial state is a consequence of the low gas fractions within the subclusters, caused by the accretion of gas onto the stars and the accretion-induced shrinkage of the subclusters. Because the subclusters are gas-poor, up to a length scale of 0.1--0.2~pc at the end of the simulation, they are only weakly affected by gas expulsion. The fraction of subclusters that reaches the high density required to evolve to a gas-poor state increases with the density of the star-forming region. We extend this argument to star cluster scales, and suggest that the absence of gas indicates that the early disruption of star clusters due to gas expulsion (infant mortality) plays a smaller role than anticipated, and is potentially restricted to star-forming regions with low ambient gas densities. We propose that in {\it dense} star-forming regions, the tidal shocking of young star clusters by the surrounding gas clouds could be responsible for the early disruption. This `cruel cradle effect' would work in addition to disruption by gas expulsion. We suggest possible methods to quantify the relative contributions of both mechanisms.
\end{abstract}


\section{Introduction} \label{sec:intro}
Over the past years, the implications of clustered star formation have touched a range of astrophysical disciplines, from the scales of the star formation process itself \citep[see the review by][]{mckee07} to the fundamental properties of young star clusters \citep[e.g.][]{mcmillan07,allison09,moeckel09}, or possibly even the global stellar mass assembly of galaxies \citep[see e.g.][]{pflamm07,bastian10}. While it seems evident that most stars form in a clustered setting \citep[e.g.][]{parker07}, estimations of the exact fraction are hampered by the substantial dissociation of stellar structure that occurs during (but is not necessarily related to) the transition from the gas-embedded phase to classical, gas-poor star clusters \citep{lada03,portegieszwart10}. The traditional interpretation that {most, if not all stars form in clusters, with} gas expulsion leading to their early disruption \citep[`infant mortality', see][]{lada03,bastian06b,goodwin06} has recently been challenged by observational studies suggesting that stars form with a continuous distribution of densities, of which only the {high-density} tail eventually leads to bound stellar clusters \citep{bressert10,gieles11}.

Current advancements in numerical calculations of turbulent fragmentation in star-forming regions enable the study of clustered star formation in increasing detail \citep[e.g.][]{bonnell98,klessen00,bate03,bonnell08}. However, theoretical investigations of the response of stellar structure to gas expulsion are still largely based on the assumption of {either a static gas potential\footnote{{Except for a normalisation of the gas potential that decreases with time when the gas is expelled.}} or} initial equilibrium between the stars and gas \citep[e.g.][]{tutukov78,adams00,geyer01,boily03,boily03b,baumgardt07,parmentier08}, which need not apply to star-forming regions in nature. A more realistic setting was recently explored by \citet{offner09}, who find that the velocity dispersions of the stars in hydrodynamic simulations of star formation are smaller than that of the gas by about a factor of~5, suggesting that the assumption of equilibrium between both components indeed does not hold. The response of star clusters to gas expulsion has also been investigated by \citet{moeckel10}, who consider $N$-body simulations of star clusters using initial conditions from hydrodynamic simulations, and by \citet{moeckel11}, who address the dynamical evolution of star clusters under the condition of ongoing gas accretion. They propose that the disruptive effect of gas expulsion is limited by the way in which gas and stars are redistributed by the accretion-induced shrinkage of clusters.

The hydrodynamical calculations of \citet{bonnell08} cover the hierarchical formation of several stellar (sub)clusters, which have been identified and analysed by \citet{maschberger10}. The simulation is very suitable for investigating the properties of the (sub)cluster population due to the relatively large {range of mass scales of the modelled structure} (see Sect.~\ref{sec:sims}). In this paper, we analyse the simulations reported in \citet{bonnell08} to probe the response of gas-embedded stellar structure to gas expulsion. We consider the dynamical state of the stars while ignoring the gas, which is equivalent to observing the stellar structure under the condition of instantaneous gas expulsion at any time in the simulation.

This paper starts with a discussion of the setup of the simulations, the subcluster identification algorithm, and the characteristics of the stellar structure in Sect.~\ref{sec:sims}. An analysis of the dynamical state of the subclusters follows in Sect.~\ref{sec:results}, which covers quantities such as the virial ratio, bound mass fraction and gas content. The response of the clusters to gas expulsion and an extension to the length scales of actual star clusters is considered in Sect.~\ref{sec:exp}. The paper is concluded with a summary and an outlook in Sect.~\ref{sec:summ}, where we discuss the possible dependence of the results on the initial conditions of the simulations and the input physics, and suggest ways in which our analysis could be improved and extended.

\section{Simulations and cluster selection} \label{sec:sims}
In this work, we analyse the hydrodynamical/$N$-body simulations of \citet{bonnell03} and \citet{bonnell08}, extending the analysis of \citet{maschberger10} and \citet{maschberger11b}. These smoothed particle hydrodynamics (SPH) simulations follow the evolution of a initially marginally unbound, homogeneous gas sphere of $10^3~\msun$ with a diameter of $1$~pc \citep{bonnell03}, and a cylinder of $3\times10$~pc that contains $10^4~\msun$ gas, bound in the upper part and unbound in the lower \citep{bonnell08}. Initial turbulent motions are modelled with an initially divergence-free random Gaussian velocity field with a power spectrum $P(k) \propto k^{-4}$. {The gas is kept at a temperature of 10~K, staying isothermal throughout the $10^3~\msun$ simulation. In the $10^4~\msun$ simulation, the gas follows a modified Larson-type equation of state \citep{larson05} comprised of three barotropic equations of state:
\begin{equation}
\label{eq:eos}
P \propto \rho^{\gamma} ,
\end{equation}
with $P$ the pressure, $\rho$ the density, and where
\begin{equation}
\begin{array}{rlrr}
\gamma  &=&  0.75  ;& \hfill \rho \le \rho_1 \\
\gamma  &=&  1.0  ;& \rho_1 < \rho  \le \rho_2 \\
\gamma  &=&  1.4  ;& \hfill \rho_2 <\rho \le \rho_3 \\
\gamma  &=&  1.0  ;& \hfill \rho > \rho_3, \\
\end{array}
\end{equation}
and $\rho_1= 5.5 \times 10^{-19}~{\rm g~cm}^{-3} , \rho_2=5.5 \times 10^{-15}~{\rm g~cm}^{-3} , \rho_3=2 \times 10^{-13}~{\rm g~cm}^{-3}$. {For reference, a density of $1~\msun~({\rm 0.01~pc})^{-3}$ equals $1.6\times 10^{-17}~{\rm g}~{\rm cm}^{-3}$.}}

Star formation is modelled via sink particles, which are formed if the densest gas particle and its $\sim 50$ neighbours are gravitationally bound (the critical density {for sink particle formation} is $1.5 \times 10^{-15}~\mathrm{g}~\mathrm{cm}^{-3}$ for the $10^3~\msun$ simulation, and $6.8 \times 10^{-14}~\mathrm{g}~\mathrm{cm}^{-3}$ for the $10^4~\msun$ simulation, see \citealt{bonnell08} for details). The mass resolutions of the sink particles are $\sim 0.1~\msun$ and $0.0167~\msun$, respectively. Accretion onto sink particles occurs if SPH particles move within the sink radius (200~AU for both simulations) and are gravitationally bound, or if SPH particles move within the accretion radius (40~AU for both simulations). Gravitational forces between sink particles are softened at smoothing lengths of 160~AU ($10^3~\msun$) and 40~AU ($10^4~\msun$).

Under the influence of gravity, the initially smooth gas distributions quickly form filaments in which the sink particles are formed. The sink particles themselves group together in subclusters that merge into larger structures, leading to the formation of one `star cluster' in the $10^3~\msun$ simulation and about three `star clusters' in the $10^4~\msun$ simulation. Throughout this paper, we will focus on the $10^4~\msun$ simulation, which contains about ten times more subclusters than the $10^3~\msun$ simulation, and therefore enables us to consider a {\it population} of subclusters rather than a select set of examples. We also ran our analyses for the $10^3~\msun$ simulation, which gave results that are consistent with those from the $10^4~\msun$ simulation.

For the identification of the subclusters, we employ a minimum spanning tree (MST) based clustering technique, which has been used in the context of young star forming regions \citep[e.g.][]{maschberger10,kirk11}. The MST, which has the advantageous property of not imposing any geometrical symmetry on the data set, has also been used to quantify the amount of substructure \citep[e.g.][]{cartwright04,schmeja08,maschberger10} and mass segregation (e.g. \citealt{allison09,allison09a,maschberger10,parker11} -- see \citealt{moeckel09} and \citealt{maschberger11b} for alternative methods) in star forming regions. The MST is a concept from graph theory, which represents the unique connection of all points of a data set, so that there are no closed loops (a `tree'), and so that the total length of all edges between the points is minimal. Typically, two separated groups of points are connected with one single, long edge, whereas the points within the groups have short edges. By simply removing edges that are longer than a chosen break distance the tree can be split in subtrees, which connect the points of the subclusters in the data set \citep[further information on MST based clustering techniques can be found in][]{zahn71}. To avoid spurious detections, we require that a subcluster contains a minimum number of 12 stars. 

\begin{figure}
\resizebox{8cm}{!}{\includegraphics{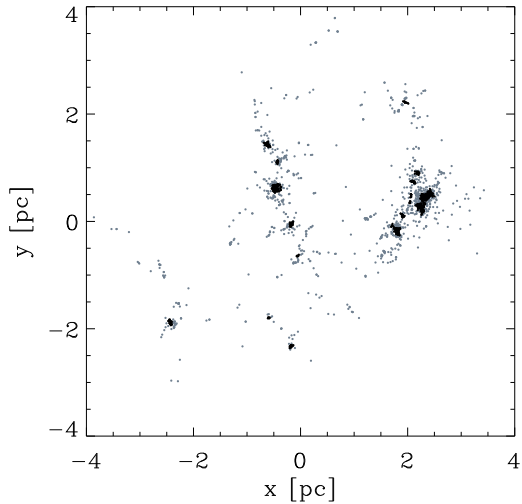}}
\caption[]{\label{fig:xy}\sf
Spatial distribution of sink particles that are present at the end of the $10^4~\msun$ simulation ($t=0.641$~Myr), projected on the x-y plane. Black particles constitute subclusters, and dark grey particles belong to the field population. {Since the spatial extent of the simulation in the z-direction is larger than in the x-y plane, some of the apparent clustering is the result of the projection.}
       }
\end{figure}
The MST technique utilises one free parameter, which is the break distance. Automated methods to determine the break distance are ill-suited for the analysis of the simulations due to the highly varying properties of the stellar distribution. We therefore choose a break distance of $d_{\rm break}=0.035~\mathrm{pc}$, which gives subclusters that are comparable to those identified by the human eye, {although they do not include the stellar haloes surrounding them}. This break distance is larger than the choice of \citet{maschberger10}, but leads to comparable clusters because we analyse the stellar structure in three spatial dimensions and not in projection. Because the choice of a single break distance could introduce an artificial length scale into the analysis \citep{bastian07}, we have also performed our calculations for a set of other break distances in the range $d_{\rm break}=0.020$--0.100~pc, of which the results are used when discussing the implications and applicability of our findings in Sects.~\ref{sec:exp} and~\ref{sec:summ}.

An example of the results of the subcluster identification method can be seen in Fig. \ref{fig:xy}, which shows the spatial distribution of sink particles and subclusters at the end of the $10^4~\msun$ simulation {at $t=0.641$~Myr, some $0.3$~Myr after the onset of star formation}. At this point in the simulation, after one free-fall time, about 60\% of all stellar mass is constituted by subclusters. The spatial distribution shown in Fig.~\ref{fig:xy} is the result of a complex tree of hierarchical merging of small subclusters \citep{maschberger10}. This process is still ongoing at the end of the simulation, which is illustrated by the close proximity of the subclusters towards the right in the plane of Fig.~\ref{fig:xy}.

\begin{figure}
\resizebox{8cm}{!}{\includegraphics{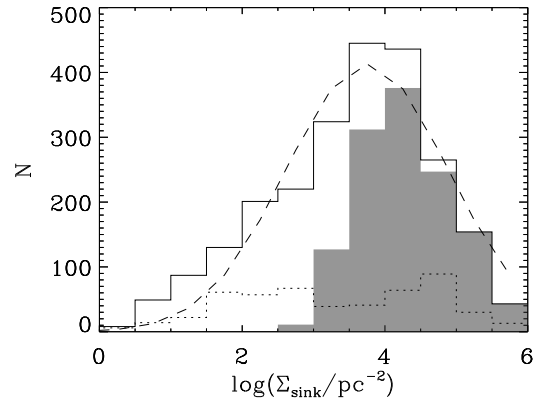}}
\caption[]{\label{fig:surfdens}\sf
Surface density distributions of sink particles. The solid histogram includes all sink particles present at the end of the simulation ($t=0.641$~Myr), with the shaded histogram marking the subset of particles that belong to subclusters. The dashed curve represents a lognormal fit to the distribution with a peak at $\log{(\Sigma_{\rm sink})}=3.75$ and a dispersion of $\sigma_{\log{\Sigma}}=1.13$. For comparison, the dotted histogram denotes the surface density distribution of sink particles at $t=0.442$~Myr.
       }
\end{figure}
\begin{figure}
\resizebox{8cm}{!}{\includegraphics{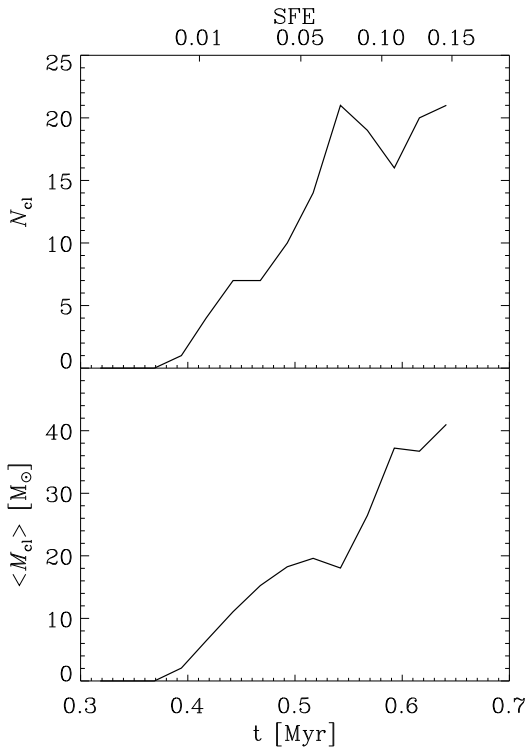}}
\caption[]{\label{fig:meanmcl}\sf
{\it Top}: time evolution of the number of subclusters. {\it Bottom}: time evolution of the mean subcluster mass. The star formation efficiency (SFE) of the entire simulation is indicated along the top axis.
       }
\end{figure}

Following \citet{casertano85}, we use the projected distance to the $N$th nearest neighbour to determine the surface density distribution of the sink particles. For a rank number $N$, the local surface density at the locations of each of the sink particles is $\Sigma_{\rm sink}=(N-1)/(\pi D_N^2)$, with $D_N$ the projected distance to the $N$th nearest neighbour {in the x-y plane}. The resulting distribution of surface densities for $N=7$ is shown in Fig.~\ref{fig:surfdens}, which includes all sink particles at the end of the simulation ($t=0.641$~Myr), as well as those from a snapshot at $t=0.442$~Myr, not too long after the onset of star formation (also see Fig.~\ref{fig:meanmcl}). The difference between the surface density distributions at both times shows that the stellar structure in the simulation typically evolves towards higher densities, even though the density {\it range} spanned by the sink particles does not change much. As should be expected, the high end of the surface density distribution is occupied by the sink particles that are residing in subclusters (shaded area), reaching densities of more than $10^5$ stars per pc$^2$. These surface densities are several orders of magnitude higher than those observed {in nearby ($<500$~pc) star-forming regions} by \citet{bressert10}, which is not surprising for two reasons. Firstly, crowding obstructs the observation of the densest parts of star-forming regions, which are therefore not included in their sample. Secondly, the high densities that are achieved in the simulation are likely the result of the initial conditions, {with a mean initial gas mass surface density of over $10^3$~\msun~pc$^{-2}$ in the x-y plane}. Nonetheless, the surface densities do compare well to the high-density region in the Orion Nebula cluster (ONC), {of which the central surface density coincides with the peak of the distribution in Fig.~\ref{fig:surfdens}} \citep{hillenbrand98}, although the system under consideration here is about one order of magnitude younger than the ONC.

The subcluster assembly history is considered in Fig.~\ref{fig:meanmcl}, which shows how the number of subclusters and the mean subcluster mass evolve as a function of time and the star formation efficiency (SFE, the fraction of the gas that has been converted to stars) {of the entire simulation} (see \citealt{bonnell11} for a detailed spatial analysis of the SFE). The number of subclusters initially increases until it reaches a maximum, which occurs when the formation of new concentrations of sink particles is neutralised by the hierarchical merging of other subclusters. This is nicely illustrated by the mean {stellar} subcluster mass, which steeply increases  around $t=0.55$--0.60~Myr, when the merging of new-formed subclusters causes the number of clusters to decrease. The mean mass keeps increasing until the end of the simulation due to the ongoing accretion of gas and small subclusters onto {sink particles and the merging of  small subclusters with a few massive ones}. This mass increase would eventually be halted on time scales much longer than the duration of the simulation, when the available gas reservoir is depleted or, more likely, when the further inflow of gas {onto the subclusters} is obstructed by feedback from supernovae and stellar winds (neither of which are included in the simulation).

\section{Dynamical state of stellar subclusters} \label{sec:results}
Whether or not a gas-embedded stellar structure survives gas expulsion depends on its dynamical state. Excluding the gas from the dynamical analysis is equivalent to observing the stellar structure at the moment of instantaneous gas removal. This represents the extreme case of gas expulsion, since a more gradual expulsion could allow a subcluster to (adiabatically) respond to the potential change and thereby retain a larger number of stars. As a result, the analysis of the dynamical state of solely the stellar component in simulations of star formation provides a lower limit to the retention of stellar structure upon gas removal.

\subsection{Dynamical quantities} \label{sec:dyn}
We have followed the evolution of several (dynamical) properties of the identified subclusters over the course of the simulation of \citet{bonnell08}, such as the stellar mass, the {stellar} half-mass radius\footnote{The subclusters have predominantly small elongations \citep[Fig.~10]{maschberger10}, which enables the use of a half-mass radius.}, the fraction of the subcluster mass that is bound, and the virial ratio. The properties of the stellar component are supplemented with information on the gas, including the gas mass fractions within the subclusters.

The gravitational boundedness and virial ratio of subclusters are fundamental measures for their dynamical state. Both quantities are based on the potential energy and internal kinetic energy of a subcluster. For a sink particle $i$, the potential energy $V_i$ is defined as
\begin{equation}
\label{eq:epot}
V_i=-\sum_{j\neq i}\frac{Gm_i m_j}{r_{ij}} ,
\end{equation}
where $m_i$ and $m_j$ are the sink particle masses and $r_{ij}$ their mutual distance. The kinetic energy $T_i$ of a sink particle is
\begin{equation}
\label{eq:ekin}
T_i=\frac{1}{2}m_i |{v}_i-{v}_{\rm cl}|^2 ,
\end{equation}
where ${v}_i$ and ${v}_{\rm cl}$ are the respective velocity vectors of the sink particle and the centre of mass of the subcluster. A particle is gravitationally bound if $T_i+V_i<0$. We define the virial ratio $Q_{\rm vir}$ as
\begin{equation}
\label{eq:vir}
Q_{\rm vir} = -\frac{2\sum_i T_i}{\sum_i V_i} .
\end{equation}
The factor 2 reflects the correction for counting the potential energy twice for each particle pair when combining Eqs.~\ref{eq:epot} and~\ref{eq:vir}. A subcluster is in virial equilibrium if $Q_{\rm vir}\sim0.5$ and gravitationally bound if $Q_{\rm vir}<1$. {Supervirial subclusters have $Q_{\rm vir}>0.5$, while subvirial subclusters have $Q_{\rm vir}<0.5$.}

It is possible that a single, dynamically hard binary, triple or multiple system dominates the energy of a subcluster. We correct for this by searching the sink particle list for binaries\footnote{Binaries are selected by identifying a most bound partner for each sink particle. If it exists and the semimajor axis is smaller than 1000 AU, it is considered a binary.} and replacing them with a single centre of mass particle. We repeat this step two more times, thereby correcting for triples and higher-order multiple systems. During the last iteration, the kinetic and potential energies of the subclusters generally remain unchanged, which indicates that a correction for higher order multiples is not required. We quantify the effect of binaries on the observables of interest below.

In previous studies, the SFE has been identified as a key parameter which determines the survival chances of stellar clusters upon gas expulsion \citep[e.g.][]{goodwin06}. However, a more fundamental critical factor is the dynamical state of the stars when the gas is removed. The {\it effective} SFE, ${\rm eSFE}=1/2Q_{\rm vir}$, was therefore introduced as a measure for the survival probability of stellar structure at the moment of instantaneous gas expulsion \citep[e.g.][]{verschueren90,goodwin09}. If the gas and stars were in virial equilibrium before gas expulsion, the eSFE is equivalent to the actual SFE. If they were not in virial equilibrium, the eSFE no longer reflects the actual SFE. In that case, the survival chance of stellar clusters is not related to the actual SFE, but is solely determined by their dynamical state. The eSFE is naturally higher in the identified subclusters than in the simulation as a whole.

\subsection{Virial ratio} \label{sec:virial}
As star formation progresses, the population of subclusters grows in terms of its total mass. The hierarchical merging of the subclusters inhibits the increase of their number and causes it to level off towards the end of the simulation, when the formation of new subclusters is balanced by their accretion onto more massive ones (see Fig.~\ref{fig:meanmcl} and \citealt{maschberger10}). Another consequence of this hierarchical buildup is that the properties of the subcluster population as a whole is not a direct representation of the evolution of individual subclusters, but also include `emergent' properties of the population due to the interactions between the subclusters. This is also relevant when considering the mean virial ratio of the subclusters as a function of time. Individual subclusters can be formed either subvirially or supervirially with respect to the total potential, and would eventually virialise with the total potential if kept in isolation. Deviations from this trend occur when subclusters merge or accrete smaller stellar aggregates, which temporarily increases the virial ratio of the merger product due to the relative velocity of the progenitors. Another thing to keep in mind is that we only include the stars in our dynamical analysis, implying that the obtained virial ratio is always higher than its actual value by an amount that depends on the gas fraction. This affects the mean virial ratio of the population of subclusters, in which there is a continuous formation of new, gas-rich subclusters, which are {typically still supervirial and only reach $Q_{\rm vir}=0.5$ after some further evolution}.

\begin{figure}
\resizebox{8cm}{!}{\includegraphics{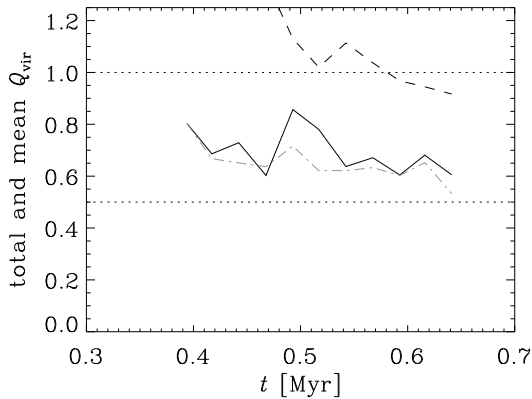}}
\caption[]{\label{fig:virial}\sf
Time evolution of the mean virial ratio of the stellar subclusters (black solid line), of the mean virial ratio weighted by subcluster mass (grey dash-dotted line), and of the total virial ratio of all sink particles in the simulation (dashed line). The horizontal dotted lines indicate the marginally gravitationally bound case ($Q_{\rm vir}<1$) and the virialised case ($Q_{\rm vir}=0.5$). After $t=0.5$~Myr, about 40--60\% of the stars resides in subclusters {for the break distance adopted here ($d_{\rm break}=0.035$~pc)}.
       }
\end{figure}
Despite the complex setting of hierarchical star formation, the evolution of the mean virial ratio can be used as a first indication of how the dynamical state of the subcluster population evolves over time. This is shown in Fig.~\ref{fig:virial}, which also includes the time evolution of the virial ratio of the entire simulation. As indicated earlier, we ignore the contribution of the gas to the gravitational potential, in order to assess the dynamical state of the stellar structure under the assumption of instantaneous gas removal. Even without accounting for the gas potential, the population of subclusters evolves to a near-virialised state on a time scale of only a few tenths of a Myr. This would suggest that the subclusters are typically gas-poor on length scales corresponding to their half-mass radii. The difference between the mean virial ratio and the mean virial ratio weighted by subcluster mass in Fig.~\ref{fig:virial} indicates that more massive subclusters are typically somewhat closer to virial equilibrium than low-mass subclusters. This is more of a trend than a relation: a simple linear regression of the virial ratio and subcluster mass $M$ yields a best fit of $Q_{\rm vir}=0.86-0.16\log{M}$, but with scatter larger than the fitted slope. Lastly, it is also shown by Fig.~\ref{fig:virial} that the entire stellar population in the simulation does not reach virial equilibrium, but does become marginally bound. This occurs because the simulation as a whole has a higher gas fraction than the subclusters (see Sect.~\ref{sec:gas}), which illustrates that the SFE depends on the location and length scale on which it is computed. The dynamical state of the entire simulation also bears some traces of the initial conditions, covering a cylinder that contains a bound upper half and an unbound lower half (see Sect.~\ref{sec:sims}). 

The virial ratios of individual subclusters do not show notable correlations with subcluster mass or half-mass radius\footnote{Except for the unbound subclusters ($Q_{\rm vir}>1$), which are generally small ($r_{\rm h}<0.04$~pc) and low-mass ($M<30~\msun$).}. Instead, they depend more strongly on the recent mass evolution of the subclusters. The virial ratio temporarily increases whenever the subcluster mass increases, be it due to the merging with other subclusters or by individual sink particles moving inside the MST break distance. When sink particles move more than a break distance away and the subcluster mass decreases, the virial ratio decreases as well. Both are natural consequences of the inclusion or exclusion of transient substructure in the identification of the subclusters. The same dependence is found when using different MST break distances to identify the subclusters: larger break distances yield more extended subclusters and consequently the mean virial ratio is higher. When set to extreme values ($d_{\rm break}>0.050$~pc), close passages of subclusters are incorrectly picked up as merger products, causing a spurious increase of the virial ratio. For the largest break distance used in our analysis ($d_{\rm break}=0.100$~pc), these fluctuations can yield mean virial ratios briefly hitting $\langle Q_{\rm vir}\rangle=1$, in clear contrast with the result from Fig.~\ref{fig:virial}. 

\begin{figure}
\resizebox{8cm}{!}{\includegraphics{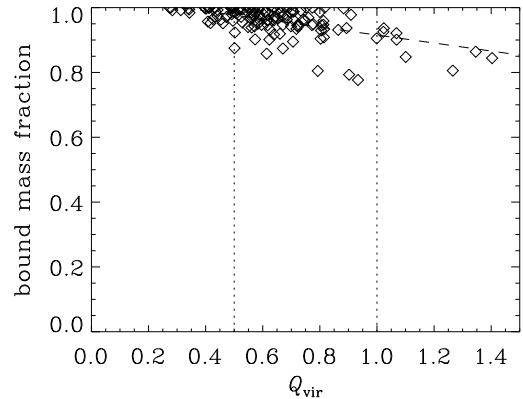}}
\caption[]{\label{fig:mbound}\sf
Dependence of the bound mass fraction {(ratio of the total mass of the bound sink particles to the subcluster mass)} on the virial ratio for the subclusters from all snapshots of the simulation. Each symbol represents a subcluster. The dashed line represents a linear fit to the data. Like in Fig.~\ref{fig:virial}, the dotted lines indicate the marginally gravitationally bound case ($Q_{\rm vir}<1$) and the virialised case ($Q_{\rm vir}=0.5$).
       }
\end{figure}
\begin{figure}
\resizebox{8cm}{!}{\includegraphics{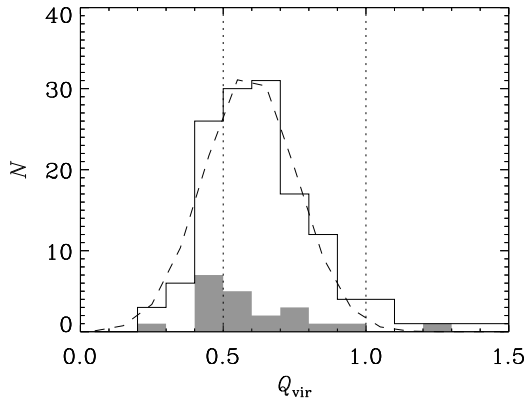}}
\caption[]{\label{fig:virial2}\sf
Histogram of the virial ratios of the subclusters from all snapshots of the simulation (solid line). The shaded histogram represents the set of subclusters from the last snapshot at $t=0.641$~Myr. The dashed line is a Gaussian fit to the data for all snapshots, with mean value $Q_{\rm vir}=0.59$ and standard deviation $\sigma_Q=0.16$. The vertical dotted lines again indicate the marginally gravitationally bound case ($Q_{\rm vir}<1$) and the virialised case ($Q_{\rm vir}=0.5$).
       }
\end{figure}

The quantity that most strongly correlates with the virial ratio is the bound mass fraction of the subclusters, i.e. the fraction of their mass that is bound even without accounting for the potential of the gas (see the definition in Sect.~\ref{sec:dyn}, a sink particle is bound if $T_i+V_i<0$.). It is shown in Fig.~\ref{fig:mbound} that subclusters with high virial ratios tend to have lower bound mass fractions, albeit with substantial scatter. This is not surprising, because the virial ratio is efficiently increased by fast, unbound sink particles {that are included by the MST algorithm but would be left out with a physically motivated identification}. For larger break distances, the correlation between bound mass fraction and virial ratio is stronger, due to the erroneous identification of kinematically hot structure as subclusters. Nonetheless, most subclusters contain only very few unbound stars, with typical bound mass fractions of $0.95$. Figure~\ref{fig:mbound} also confirms that most subclusters are close to virial equilibrium, which was already suggested by the evolution of the mean virial ratio in Fig.~\ref{fig:virial}. For further illustration, Fig.~\ref{fig:virial2} shows the distribution of the virial ratios of the subclusters from all snapshots, as well as those from the last snapshot, which are shown as the shaded region. Only 8\% of the subclusters are unbound when excluding the gas, while 25\% remains subvirial. When considering the subclusters from all snapshots, a Gaussian fit to the distribution of virial ratios gives a mean of $Q_{\rm vir}=0.59$ and a standard deviation of $\sigma_Q=0.16$. As in Fig.~\ref{fig:virial}, the gradual decrease of the mean virial ratio towards $Q_{\rm vir}=0.5$ is also visible in Fig.~\ref{fig:virial2}. A comparison of the two histograms shows that the subclusters in the last snapshot are closer to being virialised than the population of subclusters from all snapshots. These virial ratios imply that the eSFE is close to unity, i.e. the majority of subclusters will not be strongly affected by gas expulsion (see Sect.~\ref{sec:exp}).

Replacing hard binaries and higher order multiples by their centre of mass particles is essential to obtain a reliable picture of the subcluster dynamics. The disruption of the subclusters during gas expulsion is controlled by the dynamical state of the binary centres of mass rather than the binaries themselves. {Had we not corrected} for binaries or higher order multiples, the measures for the dynamical state of the subclusters would fluctuate with the orbital phase of a few tightly bound and eccentric binaries. 

The bound mass fraction of the subclusters is not strongly affected by the presence of binaries (unbound sink particles are generally single), but because binaries are in virial equilibrium\footnote{{Instantaneously}, this only holds for binaries on circular orbits. Binaries on eccentric orbits exhibit a variation of the virial ratio, with a subvirial state near apocentre and a supervirial state near pericentre. {When considering the time-averaged virial ratio, eccentric binaries are in virial equilibrium. However,} because the phase velocity is lowest near apocentre, a sampled population of randomly oriented binaries will typically be slightly subvirial.} or slightly subvirial, the mean virial ratio of the subclusters from all snapshots is decreased by 0.1--0.2 if it is not corrected for multiples. About two thirds of this difference is due to binaries, while the remaining third is accounted for by triples and quadruples. This shift of the virial ratio means that without correcting for binaries, the subclusters could be incorrectly interpreted as being slightly subvirial\footnote{As in all of Sect.~\ref{sec:virial}, this statement excludes the gravitational potential of the gas.} ($Q_{\rm vir}\sim 0.45$--0.50), and the entire simulation would be close to virialised ($Q_{\rm vir}\sim 0.60$--0.65) instead of the marginally bound state that is shown in Fig.~\ref{fig:virial}. With respect to the binary-corrected results from Fig.~\ref{fig:virial}, this rather modest difference arises because the finite {gravitational} smoothing length used in the simulation inhibits the formation of very hard binaries. Nonetheless, the correction for binaries improves the accuracy of our analysis, and therefore all results shown in this paper are corrected for binaries and higher order multiple systems.

\subsection{Gas content} \label{sec:gas}
\begin{figure}
\resizebox{8cm}{!}{\includegraphics{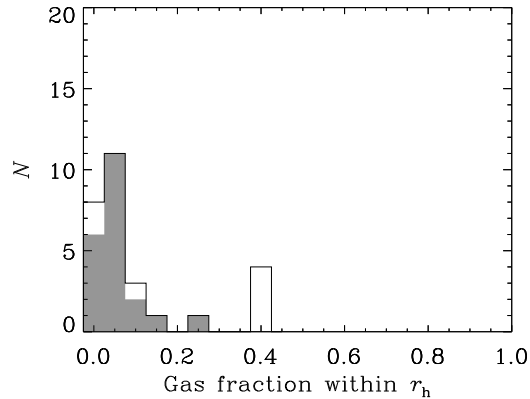}}
\caption[]{\label{fig:gas}\sf
Histogram of the gas fraction within the stellar half-mass radius $r_{\rm h}$ of each subcluster from the two snapshots at $t_1=0.442$~Myr and $t_2=0.641$~Myr (solid line). The shaded histogram only shows the gas fractions for the last snapshot at $t=0.641$~Myr.
       }
\end{figure}
The key question is {\it why} the subclusters are so close to virial equilibrium when neglecting the gas potential. An obvious answer would be that the subclusters are generally gas-poor, which would imply that they are hardly affected by the gas potential in the first place. To assess the gas potential and its time evolution, we have analysed the distribution of the gas in two snapshots of the simulation, at times $t_1=0.442$~Myr (when star formation is ongoing) and $t_2=0.641$~Myr (the last snapshot of the simulation, after one free-fall time; also see Fig.~\ref{fig:meanmcl}). For each of the identified subclusters in these snapshots, we calculate the fraction of the total mass within the {stellar} half-mass radius of the stellar distribution that is constituted by gas. The distribution of these gas fractions is shown in Fig.~\ref{fig:gas}, which confirms that the subclusters are indeed gas-poor on their typical length scales, with gas fractions of $\langle f_{\rm gas} \rangle=0$--0.2. Because the simulation does not include feedback, this means that the accretion of gas onto the sink particles can keep up with the overall gas inflow towards the subclusters. Another mechanism that naturally leads to gas-poor subclusters is their accretion-driven shrinking  \citep{bonnell98,moeckel10,moeckel11}, which increases the density contrast between the subclusters and the surrounding gas.

Gas accretion and the time evolution of the structural properties of the population of subclusters both further decrease the gas fraction as time progresses. This evolution is illustrated by comparing the data of the two snapshots in Fig.~\ref{fig:gas}, corresponding to times $t_1=0.442$~Myr and $t_2=0.641$~Myr. During the enclosed time interval, the mean gas fraction of all detected subclusters decreases by 0.63~dex, from $\langle f_{\rm gas}(t_1)\rangle=0.238$ to $\langle f_{\rm gas}(t_2)\rangle=0.056$. The mean half-mass radius of the subclusters\footnote{{The subcluster sizes are strongly correlated with the adopted MST break distance. See Sect.~\ref{sec:expand} for a discussion.}} decreases from $\langle r_{\rm h}(t_1)\rangle=0.020$~pc to $\langle r_{\rm h}(t_2)\rangle=0.013$~pc, which is a decrease of 0.18~dex. Even though the shrinking of subclusters is a second order effect caused by gas accretion, it is interesting to ask which of both mechanisms contributes most to the decrease of the gas fraction. Is it mainly driven by the increasing mean stellar density of the subclusters or by the ongoing gas accretion onto the sink particles?

\begin{figure}
\resizebox{8cm}{!}{\includegraphics{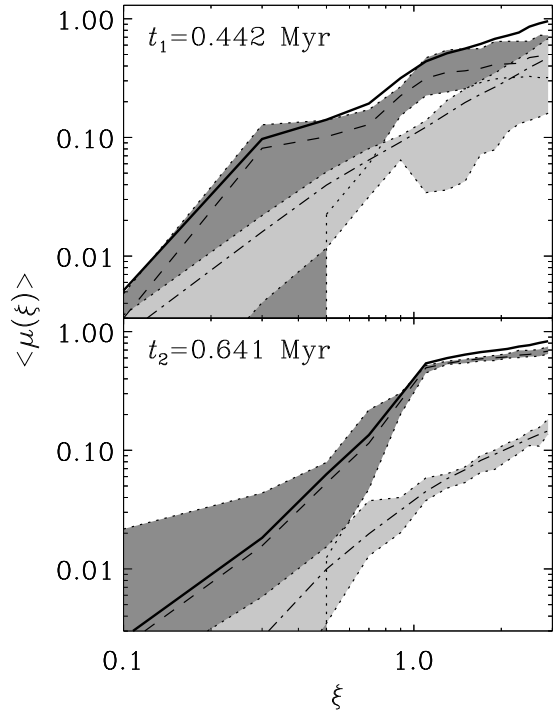}}
\caption[]{\label{fig:gas2}\sf
Subcluster mass-weighted, mean cumulative mass fractions $\langle\mu(\xi)\rangle$ (see Eq.~\ref{eq:mu}) of sink particles (dashed line), gas particles (dash-dotted line) and the sum of both (solid line), as a function of the radial distance in units of the half mass radius ($\xi\equiv r/r_{\rm h}$). {\it Top}: mean cumulative distributions for the subclusters present at $t_1=0.442$~Myr. {\it Bottom}: mean cumulative distributions for the subclusters present at $t_2=0.641$~Myr. The shaded areas enclosed by the dotted lines mark the 16th and 84th percentiles and illustrate the typical spread of the enclosed mass fractions of sink particles (dark grey) and gas (light grey).
       }
\end{figure}
To assess the relative contributions to gas depletion by accretion and subcluster shrinking, we consider the spatial distribution of the sink particles and the gas. Due to the relatively small numbers of stars in individual subclusters, it is best to examine the mean density profiles of the populations of subclusters in the two snapshots at $t_1$ and $t_2$. Such a combination of the different density profiles decreases the influence of low-number statistics on the result. In Fig.~\ref{fig:gas2}, we show the subcluster mass-weighted, mean cumulative mass distributions of gas, sink particles, and both combined. The distributions represent the enclosed mass fractions $\mu$, normalised to the sum of the subcluster mass $M_{\rm cl}$ and the enclosed gas mass within three {stellar} half-mass radii $M_{\rm gas}$:
\begin{equation}
\label{eq:mu}
\mu_i(\xi)\equiv \frac{M_i(\xi)}{M_{\rm cl}+M_{\rm gas}} ,
\end{equation}
with $i=\{{\rm stars, gas, all}\}$ and $M_i(\xi)$ the enclosed mass at $\xi\equiv r/r_{\rm h}$, which is the radial distance in units of the {stellar} half-mass radius. The mean distributions shown in Fig.~\ref{fig:gas2} are weighted by subcluster mass to emphasise those subclusters with better statistics. A first comparison of both panels in Fig.~\ref{fig:gas2} shows that the gas fraction indeed decreases between $t_1$ and $t_2$. The contribution to this decrease by subcluster shrinking can be estimated by a simple thought experiment, in which the gas distribution is kept fixed and the distribution of stellar mass is compressed by the appropriate amount. In the top panel of Fig.~\ref{fig:gas2}, the half-mass radii $\langle r_{\rm h}(t_1)\rangle$ and $\langle r_{\rm h}(t_2)\rangle$ correspond to $\xi=1$ and $\xi=0.66$, between which the enclosed gas mass differs by 0.26~dex. In other words, if the gas distribution was held fixed and the stellar distribution was shrunk appropriately, then the gas fraction within the new half-mass radius would have declined by 0.26~dex. This is a probe for the decrease of the gas fraction that is solely caused by the shrinking of the subclusters. Comparing it with the actual decrease of the mean gas fraction between $t_1$ and $t_2$ of 0.63~dex, we see that it covers about half of the decrease, with the remaining 0.37~dex covered by gas accretion itself -- not only by adding to the mass in stars, but also by decreasing the gas mass. We conclude that the evacuation of the gas due to ongoing gas accretion is about equally important for the gas depletion as the shrinking of the subclusters.

Apart from enabling a quantitative comparison of the effect of gas accretion and cluster shrinking, Fig.~\ref{fig:gas2} also demonstrates the spatial variation of the gas fraction in the subclusters. At early times, the gas is still prevalent in the outskirts of the subclusters, contributing 20--60\% of the enclosed mass at $\xi=3$. At the end of the simulation this gas has mostly vanished, leaving only a few percent of the mass within the {stellar} half-mass radius as gas, and typically 20\% at $\xi=3$. {The radial dependence of the enclosed gas fraction is qualitatively reminiscent of the model of \citet{adams00}.} It is interesting to note that the relative increase of the enclosed gas mass fraction with respect to the enclosed sink particle mass fraction only occurs at radii where the latter flattens, i.e. the subclusters only become gas-rich at radii where very little stellar mass is present. The influence of the gas on the subcluster dynamics is therefore best evaluated at radii smaller than where the flattening of $\mu_{\rm stars}$ occurs. At $t_1$, the ratio between the enclosed stellar mass and gas mass just before the flattening is about 4:1, while at $t_2$ it has increased to 11:1. This suggests that if feedback starts at a time $t>t_2$, the resulting gas expulsion will not strongly affect the subcluster dynamics, and that their virialised state (see Sect.~\ref{sec:virial}) will be largely retained.

\section{Response to gas expulsion} \label{sec:exp}
Motivated by the low gas fractions found in Sect.~\ref{sec:results}, we now address the response of the subclusters to gas expulsion in more detail.

\subsection{The expansion of subclusters} \label{sec:expand}
The long-term response of the subclusters to gas expulsion can be evaluated by once again omitting the gas from the simulations and considering only the identified stellar subclusters and their evolution towards virial equilibrium. Given a certain virial ratio and bound mass fraction, does a subcluster expand or shrink after gas removal? We combine the data from the simulations with a simple energy argument similar to \citet{hills80} to estimate how the subcluster masses and half-mass radii evolve after the expulsion of the gas. It is insightful to consider the system at two key moments.
\begin{itemize}
\item[(1)] {\it The time of instantaneous gas removal}, which is equivalent to the current system in the simulations while omitting the gas. This can be done for each snapshot, thereby providing a larger sample of subclusters than when only the last snapshot were to be considered. Of course, including subclusters from different snapshots implies a correspondingly extended range of moments of gas expulsion.
\item[(2)] {\it The time at which each subcluster attains virial equilibrium} {after removing the gas}. These times are different for each cluster per definition, but by considering the subclusters at their respective times of virialisation the long-term impact of gas removal is most clearly isolated and identified.
\end{itemize}
The evolution of the subclusters between these two moments can be quantified by evaluating the conservation of energy. For any subcluster, we can express the kinetic energy as
\begin{equation}
\label{eq:energy}
T=\frac{1}{2}M_{\rm cl}\langle v^2\rangle\equiv -VQ_{\rm vir}\approx\frac{GM_{\rm cl}^2}{2r_{\rm v}}Q_{\rm vir} ,
\end{equation}
where $r_{\rm v}$ is the virial radius, and $\langle v^2\rangle$ denotes the mean square velocity, which as a result can be written as
\begin{equation}
\label{eq:v2}
\langle v^2\rangle=Q_{\rm vir}\frac{GM_{\rm cl}}{r_{\rm v}} .
\end{equation}
The total energy at the moment of instantaneous gas removal thus becomes 
\begin{equation}
\label{eq:energy1}
E_1=(Q_{\rm vir,1}-1)\frac{GM_{\rm cl,1}^2}{2r_{\rm v,1}} ,
\end{equation}
where the relevant quantities have been marked with subscript `1' to indicate the moment of gas expulsion.

Given a deviation from virial equilibrium, a subcluster will respond by changing its radius and/or mass. As can be verified from Fig.~\ref{fig:mbound}, most subclusters contain a certain number of unbound sink particles, which were either previously retained by the gas potential, or are randomly passing the subcluster close enough to be included by the cluster identification algorithm. These unbound sink particles will escape the subcluster upon gas expulsion and take away some of the kinetic energy. We now consider a second moment in time, at which the gas-rid subcluster has reached virial equilibrium ($Q_{\rm vir,2}=0.5$) and the unbound sink particles have successfully escaped. At this time, energy conservation dictates
\begin{equation}
\label{eq:energy2}
E_1=E_2+E_{\rm esc}=-\frac{GM_{\rm cl,2}^2}{4r_{\rm v,2}}+(M_{\rm cl,1}-M_{\rm cl,2})\frac{\beta\langle v^2_2\rangle}{2} ,
\end{equation}
where $E_2$ is the total energy of the (virialised) subcluster, $E_{\rm esc}$ is the total energy of the escaped stars, and the relevant quantities have been marked with subscript `2' to indicate the moment of virialisation. The parameter $\beta$ denotes the surplus energy per unit mass of the escaped stars after they clear the potential of the subcluster, in units of its mean square velocity. The values of $\beta$ can be estimated from the simulation by computing $(T_i+V_i)/(GM_{\rm cl,2}m_i/2r_{\rm v,1})$ for each of the unbound sink particles\footnote{\label{ft:assum}The denominator holds a slightly modified form of the mean square velocity $\langle v_2^2\rangle$ (see Eq.~\ref{eq:v2}) and assumes that the virial radius does not change much between gas expulsion and virialisation (i.e. $r_{\rm v,1}\approx r_{\rm v,2}$). This is required since $r_{\rm v,2}$ is not available in the simulation. The assumption will be verified below.}. For this, we use the relation between the virial and half-mass radius corresponding to a \citet{plummer11} potential, which is given by $r_{\rm v}=1.3r_{\rm h}$ \citep[e.g.][]{heggiehut}. The escaping sink particles are the tail of an approximately Maxwellian velocity distribution of the sink particles in the subcluster, and consequently the distribution of $\beta$ declines exponentially as $f(\beta)\propto {\rm exp}(-\beta/\beta_0)$, with $\beta_0$ around unity. The mean of such a distribution equals $\beta_0$ per definition, which illustrates that unbound stars typically retain velocities similar to the mean square velocity in the subcluster after they escape.

Combining Eqs.~\ref{eq:v2} and~\ref{eq:energy2}, one obtains an expression for the evolution of the gas-rid subcluster as it approaches virial equilibrium, which relates the half-mass radii, masses, initial virial ratio and $\beta$. It is given by
\begin{equation}
\label{eq:rh}
\frac{r_{\rm h,2}}{r_{\rm h,1}}\approx\frac{r_{\rm v,2}}{r_{\rm v,1}}=\frac{1}{1-Q_{\rm vir,1}}\left[\frac{1+\beta}{2}\left(\frac{M_{\rm cl,2}}{M_{\rm cl,1}}\right)^2-\frac{\beta}{2}\frac{M_{\rm cl,2}}{M_{\rm cl,1}}\right] ,
\end{equation}
where $r_{\rm h,2}$ is the only unknown and all other variables given by the simulation. For $\beta=0$, i.e. all escaping stars are only marginally unbound, the expression returns the basic result that when unbound stars escape from a virialised system ($Q_{\rm vir,1}=0.5$), it contracts to reattain virial equilibrium\footnote{This situation, in which the naturally unbound component of a system escapes, should not be confused with the response of a virialised system to mass loss due to stellar evolution, when energy is injected into the system to unbind mass. In such a case, the surplus energy of the escaping mass is supplied by the energy injection and not by the dynamical system itself, which therefore mainly loses potential energy. This does not apply to the case under consideration in Eq.~\ref{eq:rh}, where no energy is injected and the unbound stars take away more kinetic energy than potential energy.}. Inserting typical values of $\beta=1$ (see above), $Q_{\rm vir,1}=0.59$ and $M_{\rm cl,2}/M_{\rm cl,1}=0.95$ (see Sect.~\ref{sec:virial}) in Eq.~\ref{eq:rh} yields $r_{\rm h,2}/r_{\rm h,1}=1.04$, which justifies the earlier assumption that the radius does not change much between instantaneous gas removal and virialisation (see footnote~\ref{ft:assum}). This minor expansion is driven by the slightly supervirial state of the subclusters, and inhibited by the escape of unbound stars, which have velocities larger than the escape velocity.

\begin{figure}
\resizebox{8cm}{!}{\includegraphics{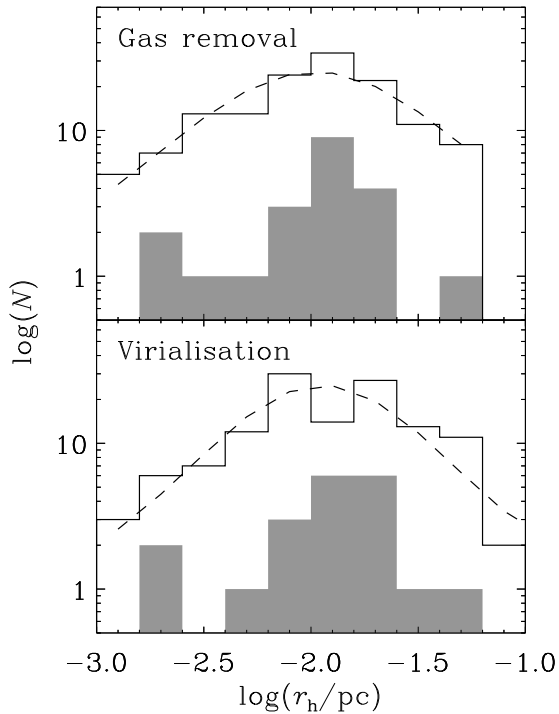}}
\caption[]{\label{fig:rh}\sf
Histogram of the {stellar} half-mass radii of the subclusters from all snapshots (solid line). The shaded histogram represents the set of subclusters from the last snapshot at $t=0.641$~Myr. {\it Top}: for the simulated radii, representing the moment of instantaneous gas removal. {\it Bottom}: for analytically computed radii, reflecting the subclusters at a later moment, when they have reached virial equilibrium. The dashed lines are lognormal fits to the data for the subclusters from all snapshots, with mean values $\log{(r_{\rm h}/{\rm pc})}=\{-1.98,-1.95\}$ and standard deviations $\sigma_{\log{r}}=\{0.73,0.61\}$ for the top and bottom panel, respectively, {with median half-mass radii $\log{(r_{\rm h}/{\rm pc})}=\{-1.96,-1.91\}$. The radii depend on the adopted break distance (see text), but the lognormal shape of the distribution persists.}
       }
\end{figure}
As discussed in Sect.~\ref{sec:virial}, the virial ratios of the subclusters give an indication of their survival fraction after gas removal. Out of all 140 subclusters identified in the simulation, only 10 have virial ratios $Q_{\rm vir}>1$ and are therefore unbound. In the last snapshot of the simulation, only one of the 21 subclusters will disperse after the removal of the gas\footnote{These unbound subclusters are typically low-mass, compact systems, and are often newly formed or have just experienced a subcluster merger.}. As a result, typically 90--95\% of all the identified subclusters survive gas expulsion. The fate of these survivors depends on whether they expand, and how their environment affects them. Expanded subclusters with lower densities are more susceptible to disruption by tidal shocks. The evolution of the half-mass radii after gas expulsion can be considered in more detail by evaluating Eq.~\ref{eq:rh} for each of the subclusters in the simulation. This enables a comparison of the distribution of half-mass radii of the current subclusters (at the moment of instantaneous gas removal) with the distribution of their half-mass radii when they have reached virial equilibrium, which is shown in Fig.~\ref{fig:rh}. The distribution of half-mass radii changes remarkably little after gas removal, as the means of the lognormal functions that are fitted to both distributions differ by 0.035~dex. This implies $r_{\rm h,2}/r_{\rm h,1}=1.08$, very similar to the earlier, simple estimate of $r_{\rm h,2}/r_{\rm h,1}=1.04$. The subclusters in the last snapshot experience roughly 1.5 times this expansion after gas removal, which is of the same order of magnitude as the expansion of the other subclusters. {As gas expulsion does not increase the cluster radii, they become no more susceptible to tidal perturbations once the gas is removed than they are at birth. However, this does not exclude any expansion of the subclusters, since mass loss due to stellar evolution and tidal perturbations would still cause them to expand (also see Sect.~\ref{sec:cce}).}

{The characteristic size of the distribution in Fig.~\ref{fig:rh} strongly correlates with the break distance of the MST. As indicated in Sect.~\ref{sec:sims}, we varied the break distance over the range $d_{\rm break}=0.020$--0.100~pc. By fitting lognormal functions to each of the obtained size distributions, we find that the corresponding mean sizes $\hat{r}_{\rm h}$ correlate with break distance as $\hat{r}_{\rm h}/{\rm pc}= (0.88\pm0.16) (d_{\rm break}/{\rm pc})^{1.33\pm0.05}$ for $0.020\leq d_{\rm break}/{\rm pc}\leq 0.100$. This demonstrates that the {\it characteristic} size of the size distribution has no physical meaning. Nonetheless, it is consistent with a lognormal distribution for all break distances, i.e. independently of the definition for the subcluster radius, and the smallest subcluster size does remain constant at about 0.005~pc, corresponding to the smallest cores in the simulation \citep{smith09}. Regardless of the adopted break distance, the smallest subclusters are always those with the lowest numbers of stars. These length scales are not very surprising, because we are not considering actual stellar clusters, but the compact (sub)systems that develop during the early formation of a cluster.

Indeed a length scale of a few 0.01~pc is much smaller than that of typical embedded clusters \citep[see e.g.][]{lada03} and of the order of the maximum size of Class~0 protostars \citep{launhardt10}. The formation of multiple stars in such a small volume could be related to the use of sink particles in the simulation, although it should be noted that small stellar clusters and multiple systems within star-forming globules are actually found on scales as small as 0.005--0.015~pc \citep{kraus08,gutermuth09,launhardt10}. Such separations are consistent with simulated low-$N$ subclusters of which the half-mass radii are dominated by a single massive star or binary. In Sect.~\ref{sec:cce}, the units of the simulation are rescaled to cover the typical mass, length and time scales of low-mass, young star clusters, causing the half-mass radii of the identified clusters to reach up to 0.3~pc.}

\subsection{The cluster formation efficiency} \label{sec:cfe}
The instantaneous gas removal discussed in this paper is an extreme form of the more gradual expulsion occurring in nature. As a result, the described weak effect of gas expulsion should be even weaker in real subclusters. It seems that gas expulsion plays a negligible role on the length scales of the compact stellar aggregates in star-forming regions. However, the regions between subclusters may still be gas-dominated, implying that feedback could prevent the further merging of subclusters and thereby inhibit their hierarchical growth.

The length scale on which the subclusters will have merged and have become gas-poor depends on the moment at which feedback starts. For the MST break distance and corresponding length scale that is used in most of this paper, the subclusters are gas-poor irrespective of time. However, there should be a break distance at which a notable time-evolution of the gas fraction appears. By comparing the subcluster gas fractions for different break distances, we find that at the end of the simulation (after one free-fall time), the subclusters have become gas-poor ($\langle f_{\rm gas}\rangle<0.1$) on a length scale of about 0.1--0.2~pc. This length scale will increase further with the number of free-fall times that are completed before the onset of {gas removal}. In turn, this increases the spatial extent over which the subclusters are allowed to merge before gas expulsion, which implies that the most massive bound structure is inversely related to the free-fall time. 

The free-fall time is related to the density as $t_{\rm ff}\propto\rho^{-1/2}$, which implies that the time of the onset of {gas removal by} feedback $t_{\rm fb}$ is associated with a density $\rho_{\rm fb}$ that has a free-fall time equal to $t_{\rm fb}$. For a given density spectrum of subclusters \citep[see e.g.][]{bressert10}, only those subclusters with densities $\rho\gg \rho_{\rm fb}\propto t_{\rm fb}^{-2}$ have the opportunity to undergo the collapse and shrinkage that we find in the simulations. The CFE  increases with the fraction of subclusters that forms in these density peaks. As subclusters merge, accrete gas and shrink, the density of the stellar structure further increases (see Sect.~\ref{sec:results}). Each free-fall time, more subclusters evolve into the density regime where $\rho\gg \rho_{\rm fb}$, also on larger length scales. This means that the length scales on which star-forming regions produce virialised stellar systems that are insensitive to gas expulsion are larger in dense sites of star formation than in sparse ones. The resulting dense clusters are also less susceptible to disruptive tidal effects from their environment, which potentially further increases their survival chances. As a result, the CFE should increase with density. Through the Schmidt-Kennicutt law \citep{schmidt59,kennicutt98b}, this suggests a relation between the CFE and the star formation rate per unit volume $\rho_{\rm SFR}$ or per unit surface area $\Sigma_{\rm SFR}$. Indeed, first observational indications for such a relation have been found by \citet{larsen00b}, \citet{larsen04b}, and recently also by \citet{goddard10}, who obtain ${\rm CFE}\propto\Sigma_{\rm SFR}^{0.24}$. A relation between the CFE and the star formation rate density would also be consistent with the high cluster formation efficiencies that are found in starburst galaxies \citep[e.g.][]{zepf99}. However, dense star-forming regions are generally also more disruptive to clustered structure due to the higher frequency and amplitude of tidal shocks \citep{lamers05a,kruijssen11}. This implies that a relation between the CFE and the ambient density would be weakened or could even be cancelled (also see Sect~\ref{sec:cce}).

\subsection{Infant mortality and the `cruel cradle effect'} \label{sec:cce}
It is often said that most stars form in stellar clusters. The concept of `infant mortality' \citep{lada03}, i.e. the rapid dispersal of stellar structure following the change of the gravitational potential due to gas expulsion, has been put forward in the literature to explain the observed rapid, mass-independent decline of the number of stellar clusters between ages of a few Myr and several tens of Myr \citep[e.g.][]{bastian06b}. Because the majority of stars has been thought to form in clusters, infant mortality was also held responsible for the low number of clusters per unit field star mass. However, recent (observational) evidence is pointing towards a picture in which star clusters are the dense end of a continuous density spectrum of star formation \citep[see Fig.~\ref{fig:surfdens} and][]{bressert10,gieles11}. This view challenges the need for infant mortality in the early disruption of stellar structure.

The results presented in this paper show that stellar substructure can evolve towards a virialised state before the gas is removed. This occurs because the dynamics of the stars and the gas are decoupled \citep[also see][]{offner09}, as the accretion of gas onto the stars together with the subcluster shrinkage can compensate the overall gas inflow onto the subclusters. In time, this causes the subclusters to become gas-poor, thereby diminishing the disruptive effect of gas expulsion. It depends on the length and mass scales on which the gas is evacuated whether this result can be extended from subclusters to actual star clusters. Towards the end of the simulation, after about 0.3~Myr of star formation, the subclusters have a mean mass of $40~\msun$ and are gas-poor on length scales of 0.1--0.2~pc. As a very crude first-order estimate, one can re-scale the units of the simulation to have the subclusters match the typical properties of young star clusters\footnote{The simulations are scale-free except for the sink particle radius and accretion radius mentioned in Sect.~\ref{sec:sims}. For the scaling used in this example, these respective radii are $10^3$~AU and 200~AU.}. By multiplying the mass unit by a factor of 25 and the length unit by a factor of 5, we re-scale the mean cluster mass to $10^3~\msun$, and the length scale on which the stellar structure will be gas-poor to 0.5--1~pc. By scaling the time unit accordingly, we see that such gas depletion is reached on a time scale of 0.8~Myr, which is of the same order as the expected $t_{\rm fb}$ due to stellar winds {and ionisation feedback}. This order-of-magnitude estimate  is of course far from conclusive, but it does show the relevance of pursuing this problem further.

If clusters reach a relatively gas-poor state before the onset of {gas removal by} feedback, the influence of gas expulsion on the dynamical state of the clusters will be smaller than previously expected. Rather than leading to the disruption of clusters (`infant mortality'), the different spatial distributions of gas and stars imply that gas expulsion could leave clusters marginally affected, unbinding only a certain fraction of their stars \citep[e.g.][]{moeckel10}. This is in clear contrast with earlier (theoretical) approaches in literature \citep[e.g.][]{boily03,boily03b,bastian06b,goodwin06,baumgardt07,parmentier08}, which assumed a model where the gas and stars are in equilibrium during gas expulsion. Clearly, this is not the case in the simulation of \citet{bonnell08}.

Because the density of the stellar structure determines whether or not gas expulsion affects the survival chances of star clusters, a continuous density spectrum of young stellar structure as in \citet{bressert10} and Fig.~\ref{fig:surfdens} naturally leads to the situation in which the dispersed part of the new-born stellar structure is affected by gas expulsion, while the other, dense and clustered part is not. However, this does not imply that the survival chances of these clusters are necessarily higher. Recent papers have argued that the disruption of star clusters due to tidal shocks from the natal environment could be substantial (\citealt{elmegreen10b}; \citealt{kruijssen11}; Kruijssen \& Bastian, in prep.). Although the disruption rate due to tidal shocks decreases with cluster density, sufficiently strong\footnote{The strength of a tidal shock corresponds to the amount of energy it injects into the cluster.} shocks would still be able to disrupt dense clusters\footnote{{Note that this early tidal disruption of stellar clusters is very different from the mechanism studied by \citet{parmentier11}, who consider the enhanced loss of stars due to the expansion of a young cluster to its tidal boundary that is induced by gas removal. The difference again lies in their assumption of equilibrium between the stars and the gas. If a young cluster has evacuated the surrounding gas and gas expulsion plays a minor role, then any gas expulsion-induced tidal stripping will be minor as well. Nonetheless, young clusters would still expand due to stellar evolutionary mass loss and the tidal shocks that are causing them to be disrupted.}} \citep{gieles06}. Such shocks could be prevalent in dense star-forming regions. As clusters move out of their primordial environment, the typical ambient gas density decreases \citep{elmegreen10b,kruijssen11}, which lessens the disruptive effect of tidal shocks. Observationally, this mechanism affects the star cluster population in a way that is very similar to infant mortality: the fact that young clusters are more efficiently disrupted than older clusters gives rise to a strong decline of the number of clusters with age at young ages. This decline acts on the age scale corresponding to the time it takes to migrate out of the star-forming region. Rather than being an internal effect, like infant mortality is, the primordial disruption by tidal shocks is driven by the state of the environment in which the clusters are born. We will therefore refer to this mechanism as the `cruel cradle effect'.

It will be interesting to quantify what the relative contributions of gas expulsion and the cruel cradle effect are to the early disruption of young stellar clusters. It is possible that both effects coexist, and that the relative importance changes with the environment. It was explained in Sect.~\ref{sec:cfe} that the fraction of clusters that is affected by gas expulsion decreases with the ambient density of the star-forming region. The cruel cradle effect shows the opposite dependence, as the disruptive effect of tidal shocks increases with the ambient gas density. It would therefore not be unlikely that gas expulsion and the cruel cradle effect each dominate a different side of the gas density spectrum of star-forming regions. Their relative strength would then determine the relation between the CFE and the ambient gas density.

\section{Summary and outlook} \label{sec:summ}
In this paper, we have assessed the dynamical state of stellar structure in star-forming regions and its response to gas expulsion by analysing the properties of the stellar structure in the SPH/sink particle simulations of \citet{bonnell03,bonnell08}. Subclusters have been identified using a minimum spanning tree algorithm \citep[MST, following][]{maschberger10}, and binaries have been replaced by their centre-of-mass particles when computing the global dynamical properties of the subclusters. We have also discussed the long-term implications of gas expulsion for the properties of star cluster populations. The main results are as follows.
\begin{itemize}
\item[(1)] The surface density distribution of sink particles follows an approximately lognormal distribution similar to that observed by \citet{bressert10}. However, the surface density corresponding to the peak of the distribution is several orders of magnitudes higher than the observed one, because the subclusters considered in our study are part of a region that would represent only one or two clusters in the observations. The high-density end of the distribution is occupied by sink particles belonging to the subclusters that are identified with the MST.
\item[(2)] When excluding the potential of the gas from the dynamical analysis and only considering the sink particles, we find that the simulation as a whole becomes marginally bound after one free-fall time, and the population of individual subclusters is close to virial equilibrium. The mean value of a Gaussian fit to the distribution of virial ratios from all snapshots is $Q_{\rm vir}=0.59$, where virial equilibrium would imply $Q_{\rm vir}=0.5$. The mean virial ratio of the population slowly decreases with time, from $Q_{\rm vir}=0.70$--0.80 early on to $Q_{\rm vir}=0.55$--0.60 towards the end of the simulation.
\item[(3)] The virialisation of the subclusters occurs due to their low gas fractions. We consider the spatial distributions of gas and sink particles at two characteristic moments during the simulation ($t_1=0.442$~Myr and $t_2=0.641$~Myr, reflecting the system early on and after one free-fall time), and find that the mean gas fraction within the {stellar} half-mass radii ($r_{\rm h}\sim 0.01$~pc) of the subclusters decreases by 0.63~dex during the enclosed time interval, from $\langle f_{\rm gas}(t_1)\rangle=0.238$ to $\langle f_{\rm gas}(t_2)\rangle=0.056$. By comparing the density profiles of gas and sink particles, we conclude that this decrease is caused by gas accretion and subcluster shrinkage to approximately the same degree.
\item[(4)] Because the subclusters are relatively gas-poor, they are only weakly affected by gas expulsion and the subsequent evolution towards virial equilibrium. {According to our analytical estimate,} they expand by an average factor of 1.08 after gas removal. The length scale on which the subclusters are gas-poor ($\langle f_{\rm gas}\rangle<0.1$) is about 0.1~pc at the end of the simulation. By scaling up the units of the simulation from subcluster to star cluster scales, we find that after about 0.8~Myr of star formation, star clusters with a mean mass of $10^3~\msun$ would be gas-poor on a length scale of 0.5--1~pc.
\item[(5)] Only those (sub)clusters with densities much larger than the density that is associated with a free-fall time {equal to the gas expulsion time} can exhibit the shrinkage and accretion that causes them to become gas poor. The fraction of clusters that reaches the required density to become insensitive to gas expulsion before the onset of {gas removal} therefore increases with ambient gas density. This suggests a relation between the cluster formation efficiency (CFE) and the ambient gas or star formation rate density, with a larger fraction of star formation resulting in bound clusters in dense regions.
\item[(6)] A possible relation between the CFE and the ambient gas density is affected by a second mechanism. In dense regions, the survival chances of stellar structure are not determined by gas expulsion or `infant mortality', but by the disruptive effect of tidal shocks from the surrounding gas \citep{elmegreen10b,kruijssen11}. This disruption of young clusters by their environment is titled the `cruel cradle effect' and is suggested to take over the disruptive role of gas expulsion in dense star-forming regions. The strength and relative contributions of infant mortality and the cruel cradle effect as a function of ambient gas density will be the decisive factor to assess the relation between the CFE and the ambient gas density. {This needs to be quantified in future studies, because the time scale covered by the simulation is too short to include the disruption of subclusters due to the cruel cradle effect.}
\end{itemize}

Throughout the paper, we have mentioned directions in which further research should be undertaken to verify and expand our conclusions. It is essential to check to what extent these results depend on the properties of the simulations we analysed, such as their initial conditions and input physics. The three chief concerns would be whether the results are affected by (1) the turbulence spectrum and initial setup of the simulation, (2) the inclusion or exclusion of feedback {and magnetic fields}, (3) the choice of sink particle radius.
\begin{itemize}
\item[(1)] The turbulence spectrum and overall boundedness of the simulation primarily influence the time evolution of the overall star formation efficiency \citep{mckee07,dale08}, and will only impact the evolution of subclusters if the inflow of gas becomes too high to be compensated by accretion and subcluster shrinkage. Judging the relative ease at which the subclusters in the current simulation become gas-poor, it is unlikely that the situation of a saturating gas inflow would take place. However, a dynamical analysis of a larger set of simulations will be needed to obtain a conclusive picture, {also to include the formation of stars and subclusters in environments with lower densities}.
\item[(2)] Feedback from accreting sink particles would inhibit the inflow of the gas, which in turn would lead to a lower gas fraction within the subclusters. There have been several efforts in literature to quantify the effect of (positive or negative) feedback on the star formation process, which generally consider effects on the length scales of the giant molecular clouds in which the star formation takes place \citep[see e.g.][]{mckee07}. While global effects could influence the gas-poor state of the (sub)clusters, the nature of feedback is such that it evacuates the gas on {the spatial scales of the subclusters}, which therefore should not lead to a fundamentally different conclusion than made in this paper. {Magnetic fields could slow down the accretion process. If this decrease of the accretion rate would be smaller in the outskirts of the subclusters than within the stellar half-mass radius, it would increase the gas fractions and virial ratios of the stellar component. Therefore, the influence of magnetic fields on our results will need to be investigated.}
\item[(3)] If the accretion and/or sink radii of the sink particles were comparable to the typical interstellar separation, the gas-poor nature of the subclusters would be a trivial result of a high `filling factor' of the subclusters by the sink particles, as the volume where the gas could reside without being accreted would be too small to achieve a stable configuration. We have addressed this to first order by computing the accretion and sink volumes taken up by sink particles and dividing it by the enclosed volume. This was done for each sink particle while taking the nearest neighbour\footnote{{We used the particle list that was corrected for multiples, implying that bound neighbours were ignored.}}, and also by calculating a mean radial `filling factor' profile for each subcluster, analogous to Fig.~\ref{fig:gas2}. Both methods returned low filling factors, with typical values of $10^{-2}$ for the sink radius and $10^{-4}$ for the accretion radius. In other words, less than 1\% of the volume inside the subclusters is taken up by the sink particles. {In order to assess to which extent this allows us to neglect spurious accretion, we ran a set of simple test simulations with different accretion and sink particle radii. These tests show that the gas accretion rate is hardly affected by the accretion and sink radii, which validates our results. The details of the tests are given in the Appendix.}
\end{itemize}
Ideally, the next step would be to perform the same type of calculation as in \citet{bonnell08} for different initial conditions, including descriptions for radiative and mechanical feedback, {potentially accounting for magnetic fields}, and varying the sink particle radius. The dynamical analysis of such simulations would provide a good verification of our conclusions, and would improve the current understanding of the dependence on initial conditions and input physics.

The order-of-magnitude extension of our results from subcluster to actual star cluster scales should be investigated further. With the continuously improving computational facilities, it will be possible to simulate systems on the scales needed to cover the formation of star clusters. The key ingredients of such an effort will be larger particle numbers and smaller sink particle radii. Additionally, infrared {or spectroscopic} observations can be used to verify the length scales on which star-forming regions are gas-poor prior to the onset of {gas removal}. The current and upcoming generation of telescopes will provide excellent opportunities for this.

If gas expulsion indeed only weakly affects the survival chances of stellar structure, it will need to be verified in which regimes infant mortality still plays a role. In order to understand the relation between the CFE and the local environment, the relative contributions to early star cluster disruption of infant mortality and the cruel cradle effect will need to be quantified. {The kinematic signatures of both effects should differ and would therefore be measurable in the velocity maps of young disrupted clusters \citep{kruijssen11c}.} Possible ways in which this could be done observationally include searching for young clusters that are currently going through gas expulsion and mapping the radial velocities of the stars, or tracing the velocity dispersion profiles of young, gas-poor clusters in dense regions. To aid this effort, the differences between the kinematic signatures of energy injection into a star cluster by gas expulsion or tidal shocks have to be established theoretically. The combination of these approaches should provide a conclusive picture of the mechanisms that determine which fraction of star formation results in bound star clusters.

\section*{Acknowledgments}
We thank the anonymous referee for thoughtful comments. JMDK is grateful to Eli Bressert for helpful discussions and acknowledges the kind hospitality of the Institute of Astronomy in Cambridge, where a large part of this work took place. ThM acknowledges funding by \textsc{Constellation}, a European Commission FP6 Marie Curie Research Training Network. This research is supported by the Leids Kerkhoven-Bosscha Fonds (LKBF) and the Netherlands Organisation for ScientiÞc Research (NWO), grant 021.001.038.

\bibliographystyle{mn2e2}
\bibliography{mybib}

\begin{thebibliography}{63}
\expandafter\ifx\csname natexlab\endcsname\relax\def\natexlab#1{#1}\fi
\small
\bibitem[{{Adams}(2000)}]{adams00}
{Adams} F.~C., 2000, \apj, 542, 964

\bibitem[{{Allison} {et~al.}(2009{\natexlab{a}}){Allison}, {Goodwin}, {Parker},
  {de Grijs}, {Portegies Zwart}, \& {Kouwenhoven}}]{allison09}
{Allison} R.~J., {Goodwin} S.~P., {Parker} R.~J., {de Grijs} R., {Portegies
  Zwart} S.~F., {Kouwenhoven} M.~B.~N., 2009{\natexlab{a}}, \apjl, 700, L99

\bibitem[{{Allison} {et~al.}(2009{\natexlab{b}}){Allison}, {Goodwin}, {Parker},
  {Portegies Zwart}, {de Grijs}, \& {Kouwenhoven}}]{allison09a}
{Allison} R.~J., {Goodwin} S.~P., {Parker} R.~J., {Portegies Zwart} S.~F., {de
  Grijs} R., {Kouwenhoven} M.~B.~N., 2009{\natexlab{b}}, \mnras, 395, 1449

\bibitem[{{Bastian} {et~al.}(2010){Bastian}, {Covey}, \& {Meyer}}]{bastian10}
{Bastian} N., {Covey} K.~R., {Meyer} M.~R., 2010, \araa, 48, 339

\bibitem[{{Bastian} {et~al.}(2007){Bastian}, {Ercolano}, {Gieles},
  {Rosolowsky}, {Scheepmaker}, {Gutermuth}, \& {Efremov}}]{bastian07}
{Bastian} N., {Ercolano} B., {Gieles} M., {Rosolowsky} E., {Scheepmaker} R.~A.,
  {Gutermuth} R., {Efremov} Y., 2007, \mnras, 379, 1302

\bibitem[{{Bastian} \& {Goodwin}(2006)}]{bastian06b}
{Bastian} N., {Goodwin} S.~P., 2006, \mnras, 369, L9

\bibitem[{{Bate} {et~al.}(2003){Bate}, {Bonnell}, \& {Bromm}}]{bate03}
{Bate} M.~R., {Bonnell} I.~A., {Bromm} V., 2003, \mnras, 339, 577

\bibitem[{{Baumgardt} \& {Kroupa}(2007)}]{baumgardt07}
{Baumgardt} H., {Kroupa} P., 2007, \mnras, 380, 1589

\bibitem[{{Boily} \& {Kroupa}(2003{\natexlab{a}})}]{boily03}
{Boily} C.~M., {Kroupa} P., 2003{\natexlab{a}}, \mnras, 338, 665

\bibitem[{{Boily} \& {Kroupa}(2003{\natexlab{b}})}]{boily03b}
---, 2003{\natexlab{b}}, \mnras, 338, 673

\bibitem[{{Bonnell} {et~al.}(2003){Bonnell}, {Bate}, \& {Vine}}]{bonnell03}
{Bonnell} I.~A., {Bate} M.~R., {Vine} S.~G., 2003, \mnras, 343, 413

\bibitem[{{Bonnell} {et~al.}(1998){Bonnell}, {Bate}, \&
  {Zinnecker}}]{bonnell98}
{Bonnell} I.~A., {Bate} M.~R., {Zinnecker} H., 1998, \mnras, 298, 93

\bibitem[{{Bonnell} {et~al.}(2008){Bonnell}, {Clark}, \& {Bate}}]{bonnell08}
{Bonnell} I.~A., {Clark} P., {Bate} M.~R., 2008, \mnras, 389, 1556

\bibitem[{{Bonnell} {et~al.}(2011){Bonnell}, {Smith}, {Clark}, \&
  {Bate}}]{bonnell11}
{Bonnell} I.~A., {Smith} R.~J., {Clark} P.~C., {Bate} M.~R., 2011, \mnras, 410,
  2339

\bibitem[{{Bressert} {et~al.}(2010){Bressert}, {Bastian}, {Gutermuth},
  {Megeath}, {Allen}, {Evans}, {Rebull}, {Hatchell}, {Johnstone}, {Bourke},
  {Cieza}, {Harvey}, {Merin}, {Ray}, \& {Tothill}}]{bressert10}
{Bressert} E., {Bastian} N., {Gutermuth} R., {Megeath} S.~T., {Allen} L.,
  {Evans} II N.~J., {Rebull} L.~M., {Hatchell} J., {Johnstone} D., {Bourke}
  T.~L., {Cieza} L.~A., {Harvey} P.~M., {Merin} B., {Ray} T.~P., {Tothill}
  N.~F.~H., 2010, \mnras, 409, L54

\bibitem[{{Cartwright} \& {Whitworth}(2004)}]{cartwright04}
{Cartwright} A., {Whitworth} A.~P., 2004, \mnras, 348, 589

\bibitem[{{Casertano} \& {Hut}(1985)}]{casertano85}
{Casertano} S., {Hut} P., 1985, \apj, 298, 80

\bibitem[{{Dale} \& {Bonnell}(2008)}]{dale08}
{Dale} J.~E., {Bonnell} I.~A., 2008, \mnras, 391, 2

\bibitem[{{Elmegreen} \& {Hunter}(2010)}]{elmegreen10b}
{Elmegreen} B.~G., {Hunter} D.~A., 2010, \apj, 712, 604

\bibitem[{{Geyer} \& {Burkert}(2001)}]{geyer01}
{Geyer} M.~P., {Burkert} A., 2001, \mnras, 323, 988

\bibitem[{{Gieles} \& {Portegies Zwart}(2011)}]{gieles11}
{Gieles} M., {Portegies Zwart} S.~F., 2011, \mnras, 410, L6

\bibitem[{{Gieles} {et~al.}(2006){Gieles}, {Portegies Zwart}, {Baumgardt},
  {Athanassoula}, {Lamers}, {Sipior}, \& {Leenaarts}}]{gieles06}
{Gieles} M., {Portegies Zwart} S.~F., {Baumgardt} H., {Athanassoula} E.,
  {Lamers} H.~J.~G.~L.~M., {Sipior} M., {Leenaarts} J., 2006, \mnras, 371, 793

\bibitem[{{Goddard} {et~al.}(2010){Goddard}, {Bastian}, \&
  {Kennicutt}}]{goddard10}
{Goddard} Q.~E., {Bastian} N., {Kennicutt} R.~C., 2010, \mnras, 405, 857

\bibitem[{{Goodwin}(2009)}]{goodwin09}
{Goodwin} S.~P., 2009, \apss, 324, 259

\bibitem[{{Goodwin} \& {Bastian}(2006)}]{goodwin06}
{Goodwin} S.~P., {Bastian} N., 2006, \mnras, 373, 752

\bibitem[{{Gutermuth} {et~al.}(2009){Gutermuth}, {Megeath}, {Myers}, {Allen},
  {Pipher}, \& {Fazio}}]{gutermuth09}
{Gutermuth} R.~A., {Megeath} S.~T., {Myers} P.~C., {Allen} L.~E., {Pipher}
  J.~L., {Fazio} G.~G., 2009, \apjs, 184, 18

\bibitem[{{Heggie} \& {Hut}(2003)}]{heggiehut}
{Heggie} D., {Hut} P., 2003, {The Gravitational Million-Body Problem: A
  Multidisciplinary Approach to Star Cluster Dynamics}. Cambridge University
  Press, 2003, 372 pp.

\bibitem[{{Hillenbrand} \& {Hartmann}(1998)}]{hillenbrand98}
{Hillenbrand} L.~A., {Hartmann} L.~W., 1998, \apj, 492, 540

\bibitem[{{Hills}(1980)}]{hills80}
{Hills} J.~G., 1980, \apj, 235, 986

\bibitem[{{Kennicutt}(1998)}]{kennicutt98b}
{Kennicutt} Jr. R.~C., 1998, \apj, 498, 541

\bibitem[{{Kirk} \& {Myers}(2011)}]{kirk11}
{Kirk} H., {Myers} P.~C., 2011, \apj, 727, 64

\bibitem[{{Klessen} \& {Burkert}(2000)}]{klessen00}
{Klessen} R.~S., {Burkert} A., 2000, \apjs, 128, 287

\bibitem[{{Kraus} \& {Hillenbrand}(2008)}]{kraus08}
{Kraus} A.~L., {Hillenbrand} L.~A., 2008, \apjl, 686, L111

\bibitem[{{Kruijssen}(2011)}]{kruijssen11c}
{Kruijssen} J.~M.~D., 2011, in Cluster Disruption: From infant mortality to
long term survival, in Stellar Clusters and Associations - A
RIA workshop on GAIA, ed., T. Gallego, {\tt ArXiv:1107.2114}

\bibitem[{{Kruijssen} {et~al.}(2011){Kruijssen}, {Pelupessy}, {Lamers},
  {Portegies Zwart}, \& {Icke}}]{kruijssen11}
{Kruijssen} J.~M.~D., {Pelupessy} F.~I., {Lamers} H.~J.~G.~L.~M., {Portegies
  Zwart} S.~F., {Icke} V., 2011, \mnras, 414, 1339

\bibitem[{{Lada} \& {Lada}(2003)}]{lada03}
{Lada} C.~J., {Lada} E.~A., 2003, \araa, 41, 57

\bibitem[{{Lamers} {et~al.}(2005){Lamers}, {Gieles}, \& {Portegies
  Zwart}}]{lamers05a}
{Lamers} H.~J.~G.~L.~M., {Gieles} M., {Portegies Zwart} S.~F., 2005, \aap, 429,
  173

\bibitem[{{Larsen}(2004)}]{larsen04b}
{Larsen} S.~S., 2004, \aap, 416, 537

\bibitem[{{Larsen} \& {Richtler}(2000)}]{larsen00b}
{Larsen} S.~S., {Richtler} T., 2000, \aap, 354, 836

\bibitem[{{Larson}(2005)}]{larson05}
{Larson} R.~B., 2005, \mnras, 359, 211

\bibitem[{{Launhardt} {et~al.}(2010){Launhardt}, {Nutter}, {Ward-Thompson},
  {Bourke}, {Henning}, {Khanzadyan}, {Schmalzl}, {Wolf}, \&
  {Zylka}}]{launhardt10}
{Launhardt} R., {Nutter} D., {Ward-Thompson} D., {Bourke} T.~L., {Henning} T.,
  {Khanzadyan} T., {Schmalzl} M., {Wolf} S., {Zylka} R., 2010, \apjs, 188, 139

\bibitem[{{Maschberger} \& {Clarke}(2011)}]{maschberger11b}
{Maschberger} T., {Clarke} C.~J., 2011, \mnras, 416, 541

\bibitem[{{Maschberger} {et~al.}(2010){Maschberger}, {Clarke}, {Bonnell}, \&
  {Kroupa}}]{maschberger10}
{Maschberger} T., {Clarke} C.~J., {Bonnell} I.~A., {Kroupa} P., 2010, \mnras,
  404, 1061

\bibitem[{{McKee} \& {Ostriker}(2007)}]{mckee07}
{McKee} C.~F., {Ostriker} E.~C., 2007, \araa, 45, 565

\bibitem[{{McMillan} {et~al.}(2007){McMillan}, {Vesperini}, \& {Portegies
  Zwart}}]{mcmillan07}
{McMillan} S.~L.~W., {Vesperini} E., {Portegies Zwart} S.~F., 2007, \apjl, 655,
  L45

\bibitem[{{Moeckel} \& {Bate}(2010)}]{moeckel10}
{Moeckel} N., {Bate} M.~R., 2010, \mnras, 404, 721

\bibitem[{{Moeckel} \& {Bonnell}(2009)}]{moeckel09}
{Moeckel} N., {Bonnell} I.~A., 2009, \mnras, 400, 657

\bibitem[{{Moeckel} \& {Clarke}(2011)}]{moeckel11}
{Moeckel} N., {Clarke} C.~J., 2011, \mnras, 410, 2799

\bibitem[{{Offner} {et~al.}(2009){Offner}, {Hansen}, \& {Krumholz}}]{offner09}
{Offner} S.~S.~R., {Hansen} C.~E., {Krumholz} M.~R., 2009, \apjl, 704, L124

\bibitem[{{Parker} {et~al.}(2011){Parker}, {Bouvier}, {Goodwin}, {Moraux},
  {Allison}, {Guieu}, \& {G{\"u}del}}]{parker11}
{Parker} R.~J., {Bouvier} J., {Goodwin} S.~P., {Moraux} E., {Allison} R.~J.,
  {Guieu} S., {G{\"u}del} M., 2011, \mnras, 412, 2489

\bibitem[{{Parker} \& {Goodwin}(2007)}]{parker07}
{Parker} R.~J., {Goodwin} S.~P., 2007, \mnras, 380, 1271

\bibitem[{{Parmentier} {et~al.}(2008){Parmentier}, {Goodwin}, {Kroupa}, \&
  {Baumgardt}}]{parmentier08}
{Parmentier} G., {Goodwin} S.~P., {Kroupa} P., {Baumgardt} H., 2008, \apj, 678,
  347

\bibitem[{{Parmentier} \& {Kroupa}(2011)}]{parmentier11}
{Parmentier} G., {Kroupa} P., 2011, \mnras, 411, 1258

\bibitem[{{Pflamm-Altenburg} {et~al.}(2007){Pflamm-Altenburg}, {Weidner}, \&
  {Kroupa}}]{pflamm07}
{Pflamm-Altenburg} J., {Weidner} C., {Kroupa} P., 2007, \apj, 671, 1550

\bibitem[{{Plummer}(1911)}]{plummer11}
{Plummer} H.~C., 1911, \mnras, 71, 460

\bibitem[{{Portegies Zwart} {et~al.}(2010){Portegies Zwart}, {McMillan}, \&
  {Gieles}}]{portegieszwart10}
{Portegies Zwart} S.~F., {McMillan} S.~L.~W., {Gieles} M., 2010, \araa, 48, 431

\bibitem[{{Schmeja} {et~al.}(2008){Schmeja}, {Kumar}, \&
  {Ferreira}}]{schmeja08}
{Schmeja} S., {Kumar} M.~S.~N., {Ferreira} B., 2008, \mnras, 389, 1209

\bibitem[{{Schmidt}(1959)}]{schmidt59}
{Schmidt} M., 1959, \apj, 129, 243

\bibitem[{{Smith} {et~al.}(2009){Smith}, {Clark}, \& {Bonnell}}]{smith09}
{Smith} R.~J., {Clark} P.~C., {Bonnell} I.~A., 2009, \mnras, 396, 830

\bibitem[{{Tutukov}(1978)}]{tutukov78}
{Tutukov} A.~V., 1978, \aap, 70, 57

\bibitem[{{Verschueren}(1990)}]{verschueren90}
{Verschueren} W., 1990, \aap, 234, 156

\bibitem[{Zahn(1971)}]{zahn71}
Zahn C.~T., 1971, IEEE Transactions on Computers, 20, 68

\bibitem[{{Zepf} {et~al.}(1999){Zepf}, {Ashman}, {English}, {Freeman}, \&
  {Sharples}}]{zepf99}
{Zepf} S.~E., {Ashman} K.~M., {English} J., {Freeman} K.~C., {Sharples} R.~M.,
  1999, \aj, 118, 752

\end{thebibliography}

\appendix
\section{Independence of results on sink parameters}

In this appendix we verify that the resolution of the SPH simulation
is not playing an important role in the evolution of the
stellar-to-gas mass ratio of the subclusters. We accomplish this via a
series of controlled, idealised tests in which a cluster of 10 sink
particles accretes from an envelope of gas. The total mass of the
system is 1, divided equally between the sinks and the gas. The sinks
are initially of equal mass, thus each has mass 0.05. They are placed
randomly in a Plummer model of virial radius $r_{\rm sinks} = 1$, and we
use the same initial configuration of the stars in each test. The
median nearest neighbour separation of the sinks is 0.43. The gas is
likewise in a Plummer sphere spatially, although with a larger radius
than the sinks. The gas has zero initial kinetic energy and minimal
thermal support, so that the gas falls onto the sink cluster and is
accreted. We run two sets of tests, one in which the gas sphere's
virial radius is ten times $r_{\rm sinks}$, i.e. $r_{\rm gas} = 10$, and one
in which $r_{\rm gas} = 3$.

\begin{figure}
\resizebox{8cm}{!}{\includegraphics{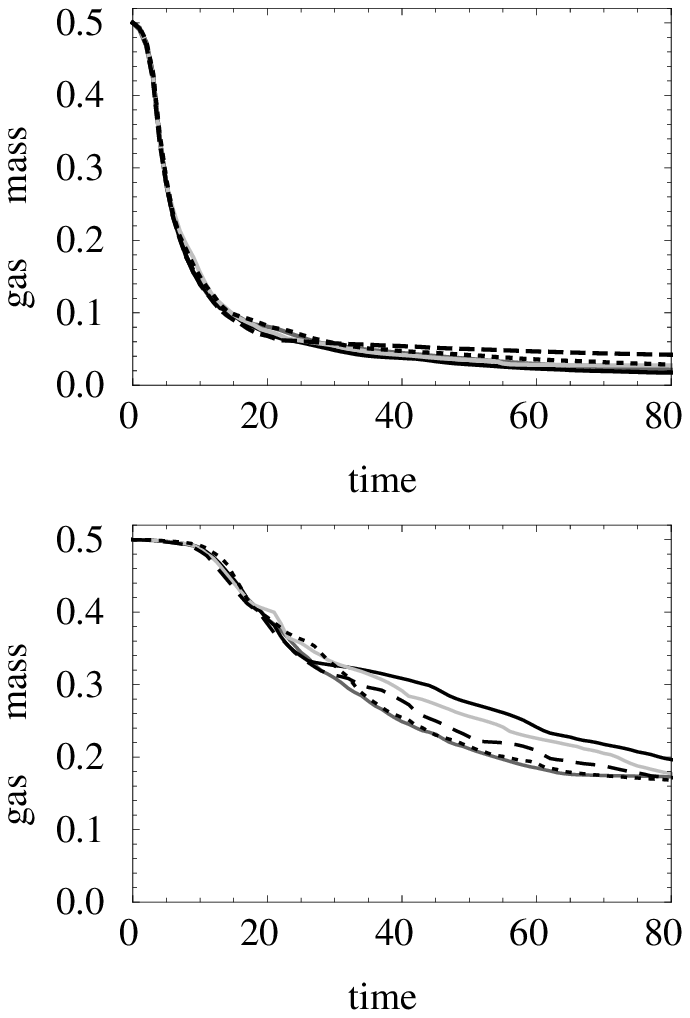}}
\caption{\sf Remaining gas mass as a function of time for our tests of
the SPH simulation sink parameters. Top panel: the runs with $r_{\rm gas}
= 3 r_{\rm sinks}$; bottom panel, the runs with $r_{\rm gas} = 10 r_{\rm sinks}$.
Grey lines have $r_{\rm acc} = 0.125$, black lines have $r_{\rm acc} = 0.0625$,
and light grey lines have $r_{\rm acc} = 0.03125$. Solid lines have
$5\times10^4$ gas particles. The dashed lines have $2\times10^5$
particles, and the dotted lines have $1.25\times10^4$ particles.
}
\label{fig:AppendixFigure}
\end{figure}
The two numerical scales we are concerned with are the accretion
radius of the sinks $r_{\rm acc}$, and the smoothing length of the gas particles.
For the sink radii, we use the set $r_{\rm acc} = \{0.125, 0.0625,
0.03125\}$. The middle value yields approximately the ratio of the
neighbour distance to the accretion radius seen in the clusters in the
simulation. The smoothing length of the gas is determined by the
number of gas particles. To roughly match the simulated value, suppose
the sinks have masses $1~\msun$. The total gas mass is then $10~\msun$, and $5\times10^4$ gas particles approximates the
resolution of the large-scale simulation. We run the 0.0625 accretion
radius cases with four times more and fewer gas particles, i.e.
$2\times10^5$ and $1.25\times10^4$.

In Fig.~\ref{fig:AppendixFigure} we show the gas mass as a function of
time for the test runs. In the top panel we show the results for the
case with $r_{\rm gas} = 3$. Time is measured dimensionlessly where we
have taken the gravitational constant $G = 1$; the crossing time of
the sink cluster is $\sim 2$. In this setup, the gas free-fall time is
$\sim 6$ and the gas accretes quickly compared to the time for the
$N$-body dynamics to dissolve the small-$N$ sink system. The
agreement between all the test runs is excellent. The bottom panel
shows the $r_{\rm gas} = 10$ cases, and the gas falls onto the sink system
over a longer time scale, with a free-fall time $\sim 35$. At early
times the agreement is quite good, with some disagreement between the
runs appearing after $t \sim 20$. We attribute this to the fact that
at this point the $N$-body dynamics of the different runs have set
the clusters on clearly divergent paths; recall that the gravitational
smoothing length of the sinks is proportional to their sink radius. By
$t = 80$ the cluster of sinks has effectively dissolved. We conclude
that gas accretion time scale is not greatly affected by the choices of
the sink radius.

\bsp
\label{lastpage}

\end{document}